\documentclass[10pt,english,a4paper,aps,superscriptaddress,prl,showkeys,showpacs,preprint]{revtex4-1}

\usepackage[T1]{fontenc}
\usepackage{geometry}
\usepackage[left]{lineno}
\usepackage{amssymb}
\usepackage{graphicx}

\makeatletter

\@ifundefined{textcolor}{}
{%
 \definecolor{BLACK}{gray}{0}
 \definecolor{WHITE}{gray}{1}
 \definecolor{RED}{rgb}{1,0,0}
 \definecolor{GREEN}{rgb}{0,1,0}
 \definecolor{BLUE}{rgb}{0,0,1}
 \definecolor{CYAN}{cmyk}{1,0,0,0}
 \definecolor{MAGENTA}{cmyk}{0,1,0,0}
 \definecolor{YELLOW}{cmyk}{0,0,1,0}
}

\usepackage{times}
\makeatother
\usepackage{babel}

\usepackage{color}

\begin{document}
\title{Controlling contagious processes \\on temporal networks via 
adaptive rewiring}

\author{Vitaly Belik} \affiliation{Institut f\"ur Theoretische Physik,
  Technische Universit\"at Berlin, Hardenbergstra{\ss}e 36, 10623
  Berlin, Germany} \affiliation{Helmholtz-Zentrum f\"ur
  Infektionsforschung, Mascheroder Weg 1, 38124 Braunschweig, Germany}

\author{Alexander Fengler} \affiliation{Institut f\"ur Theoretische
  Physik, Technische Universit\"at Berlin, Hardenbergstra{\ss}e 36,
  10623 Berlin, Germany}

\author{Florian Fiebig} \affiliation{Institut f\"ur Theoretische
  Physik, Technische Universit\"at Berlin, Hardenbergstra{\ss}e 36,
  10623 Berlin, Germany}

\author{Hartmut H. K. Lentz} \affiliation{Institute of Epidemiology,
  Friedrich-Loeffler-Institute, S\"udufer 10, 17493 Greifswald}

\author{Philipp H\"ovel} \affiliation{Institut f\"ur Theoretische
  Physik, Technische Universit\"at Berlin, Hardenbergstra{\ss}e 36,
  10623 Berlin, Germany} \affiliation{Bernstein Center for
  Computational Neuroscience Berlin, Humboldt Universit\"at zu Berlin,
  Philippstra{\ss}e 13, 10115 Berlin, Germany}

\begin{abstract}
We consider recurrent contagious processes on a time-varying
network. As a control procedure to mitigate the epidemic, we propose
an adaptive rewiring mechanism for temporary isolation of infected
nodes upon their detection. As a case study, we investigate the
network of pig trade in Germany.  Based on extensive numerical
simulations for a wide range of parameters, we demonstrate that the
adaptation mechanism leads to a significant extension of the parameter
range, for which most of the index nodes (origins of the epidemic)
lead to vanishing epidemics. Furthermore the performance of adaptation
is very heterogeneous with respect to the index node. We quantify the
success of the proposed adaptation scheme in dependence on the
infectious period and detection times. To support our findings we
propose a mean-field analytical description of the problem.
\end{abstract}


\maketitle

\section{Introduction}
\label{sec:intro}
Recently the availability of data on host mobility and contact
patterns of high resolution offers many opportunities for the design
of new tools and approaches for modeling and control of epidemic
spread~\cite{DAN13a,BAR14,STE11,SCH13a,BEL11}.  To exploit these
versatile concepts of complex networks research, hosts or their
spatial aggregations are considered as nodes and host contacts or
their relocations as edges. Very frequently the edges are not static,
but changing with time. If dynamical processes on networks possess a
characteristic time scale much faster than the time scale of the
changing edges, a static, quenched approximation of the topology may
be sufficient. However if the time scale of the process is comparable
with the time scale of the network change, a more sophisticated
concept of time-varying or temporal networks is required, because
static approximation might violate the causality
principle~\cite{CAS12,LEN13,HOL12,SCH14k,KON12a,KIV12,BAJ12,VOL09,VAL15a}.
Although for static networks a variety of surveillance and control
approaches were proposed based on various network measures such as
node degree, betweenness centrality etc.~\cite{SCH11h}, control
concepts for temporal networks are still missing. The previous studies
on this topic were devoted mostly to targeted vaccination
policies~\cite{LEE12,STA13,BUE13,LIU14,RIZ14,TAN11a} and general
controllability questions \cite{POS14, SEL14}.  Furthermore, an
adaptation of edges was proposed to mitigate the spread in static
networks \cite{GRO06b, GRO08a, SHA08a, VAN10a, YAN15} and by random
rewiring, a co-evolution of the network and the spreading was
implemented to avoid infected nodes. However, there have been no
studies combining the adaptation approach of epidemic control with
intrinsic temporal changes of the underlying network structure.

In this paper, we propose an adaptive non-targeted control mechanism
for spreading processes on temporal networks and assess its
effectiveness. We consider a deterministic recurrent contagious
dynamics similar to a susceptible-infected-susceptible (SIS) model. In
our model, susceptible nodes, after contact with an infected node,
become infected and after a fixed time $\kappa$ they become again
susceptible to the disease. We assume that the nodes are screened for
infection, but the information whether or not a node is infected is
available only after some detection time $\delta$. One can interpret
the detection time $\delta$ as a time required to reliably detect the
disease (also called window period~\cite{BRO04}) or a time the disease
needs to manifest itself (incubation time) and be diagnosed. After the
detection we apply adaptation rules, rewiring our system in a way to
avoid edges emanating from the detected infected nodes. This results
effectively in a temporary quarantine of the infected nodes..

We pursue the question, if the interplay between the intrinsic
dynamics of a temporal network and adaptation rules leads to a
substantial improvement of the disease mitigation. In contrast to a
range of studies considering targeted intervention measures, we apply
control measures to all infected nodes, after they are detected as
those. Our approach can be easily supplemented by targeted
interventions as well, where only some fraction of specific nodes is
controled.

This paper is structured as follows: First, we introduce the empirical
dataset and outline our approach to adaptive epidemic control. Then,
we present results on the mitigation strategy and discuss their
implications. Finally, we summarize our findings and provide an
outlook on further research directions.

\section{Methods and Dataset}
\label{sec:methods}

The empirical temporal network investigated in our study is extracted
from the database on pig trade in Germany~\textit{HI-Tier} (See also
Data Accessibility Section).  We use an excerpt from this animal (pig)
trade network with 15,569 agricultural premises (nodes) over a period
of observation of 2 years (daily resolution), which consists of
748,430 trade events (links). All premises have been anonymized for
the study. On average day (except Sunday) the network contains 1220
nodes with 1141 edges as depicted in Fig.~2(a) (Supplementary
Material).  We observe a non-uniform activity with respect to the day
of the week (see Supplementary Material, Fig.~1): on average, we find
1346 nodes with 1258 edges on a working day and 932 or 43 nodes with
844 or 27 edges on Saturday and Sunday, respectively. This reduced
activity on the weekend is clearly visible in Fig.~2(a) (Supplementary
Material). The trade flow is directed: from source nodes (piglet
producers) to sink nodes (slaughter houses). There are 291 sinks,
which only have in-coming links, but no out-going links, and 5,504
sources with only in-coming links. For fuhrer details on basic network
characteristics, see Tab.~1 in Supplementary Material.

The considered network also possesses a strong heterogeneity with
respect to the size of out-components as shown in Fig~2(b)
(Supplementary Material).  The out-component of a node is defined as
the number of nodes that could be infected in a worst-case SI
(susceptible-infected) epidemic scenario with an outbreak originating
from that particular node upon its first occurrence. For this worst
case, we assume the infinite infectious period and consider an SI
epidemic following the directed, temporal links during the whole
observation time.  The distribution of the out-components peaks around
6,000 nodes, which corresponds to 40\% of the network. We also find
many cases, where an outbreak immediately stops and the length of the
respective epidemic path is short. Among these nodes are the
above-mentioned sink nodes.

Therefore, it can be expected that the prevalence, i.e. the total
number of infected nodes at a given point in time, strongly depends on
the outbreak origin, and a surveillance strategy that randomly
selects nodes for screening will not be effective. In addition, for
finite infectious periods, the day of first infection is important as
has been shown in Ref.~\cite{KON12a}.

\begin{figure}[t!]
\includegraphics[width=0.6\textwidth]{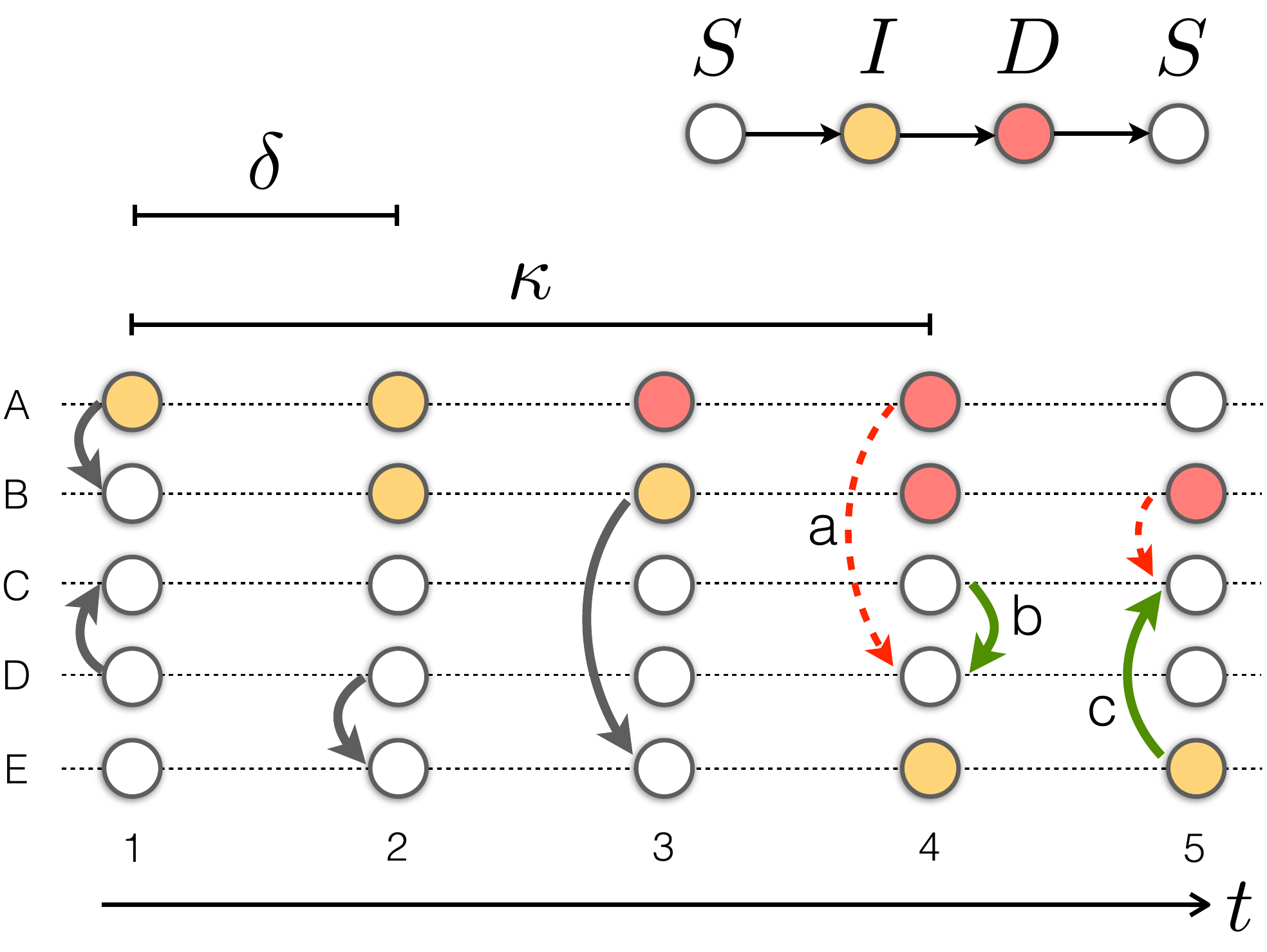}
\caption{(Color online) Infection model and adaptation
  mechanism. Nodes are arranged vertically with links between them for
  each day.
Nodes can be in three states: susceptible, $S$ (empty circles),
undetected infected, $I$ (yellow), and detected infected, $D$
(red). After $\delta$ days, infected nodes are detected and for
$t\in[t^{\star}+\delta,t^{\star}+\kappa]$ -- with $t^{\star}$ being
the time of infection -- all out-going links from the detected nodes
(red) are randomly rewired to other susceptible or undetected nodes as
starting point, e.g.~the link \textbf{a} is destroyed and instead the
link \textbf{b} is created. Note that a newly created links might
spread the disease as well. E.g. on day 5 the green arrow might point
from the undetected infected node E to the node C (link \textbf{c})
spreading the disease.
After $\kappa$ days, infected nodes become susceptible
again. Parameters: $\delta=2$~d, $\kappa=4$~d.
\label{fig: adaptation}}
\end{figure}

In our simulations on this real-world temporal network, we consider a
deterministic recurrent epidemic of an SIS type. Note that SIS
epidemics on temporal networks without adaptation have been
investigated in Ref.~\cite{ROC13}.  The spreading process is
deterministic in the following sense: every time a trade event from an
infected to a susceptible premise takes place, the susceptible one
becomes infected with probability 1. In other words, we consider
diseases with high infectiousness and thus a worst-case scenario for a
spreading process. More precisely, nodes can be susceptible ($S$),
infected undetected ($I$), or infected detected ($D$) as depicted in
Fig~\ref{fig: adaptation}. After detection time $\delta$, infected
nodes are detected and according to our control strategy, all of their
out-going links are rewired to start at other susceptible or infected,
but not yet detected nodes chosen at random. Thus, we isolate the
detected infectious nodes or place them under quarantine.  The total
period of infection is denoted by $\kappa\geq\delta$.  Note that
nodes, which are infected, but not yet detected, take part in the
rewiring and thus, eventually increase the risk of receiving
nodes. See, for instance, the newly formed (green) link from node E to
C at $t=5$~d in Fig.~\ref{fig: adaptation}.  The proposed adaptation
scheme can be easily implemented for SIR models, where the infected
nodes become immune, that is, recovered, after the infection
period. This would, however, considerably reduce the pool of available
nodes over the course of the available observation period.
%
\section{Results and discussions}
\label{sec:results}
\begin{figure}[t!]
\includegraphics[width=0.6\textwidth]{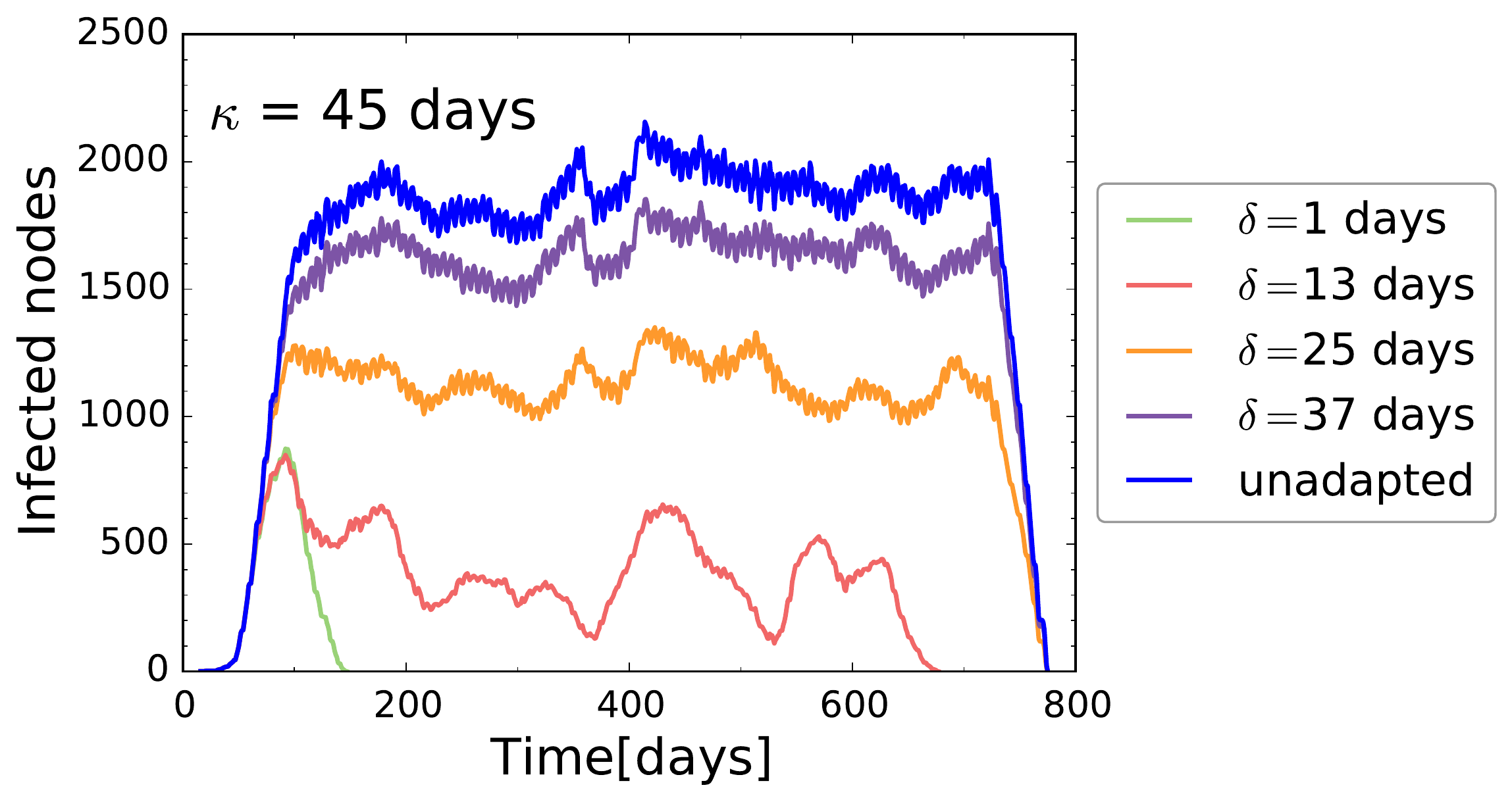}
\caption{(Color online) Typical time course of an epidemic. Prevalence
  (number of infected nodes, daily resolution) for a fixed infectious
  period $\kappa=30$~d and different detection times $\delta$. Whereas
  unadapted epidemic with $\delta=\kappa$ (blue curve), adapted
  epidemics with $\delta=37$~d (purple), $\delta=25$~d (orange) are
  persistent, adapted epidemics with $\delta=13$~d (red) and
  $\delta=1$~d (green) die out before the end of the observation.}
\label{fig:prevalence}
\end{figure}

In this section, we discuss the main findings based on the model and
data described above.  A typical evolution of the number of infected
nodes (prevalence) is shown in Fig.~\ref{fig:prevalence} for an
arbitrary starting node and the infectious period $\kappa=45$~d. For a
free-running sustained disease, i.e. without network adaptation or
other measures of mitigation ($\kappa=\delta$), one can observe that
the prevalence fluctuates around a constant endemic level after a
short transient period as shown by the blue curve.
The effect of network adaptation manifests itself either in a
reduction of this endemic prevalence level (orange curve in
Fig.~\ref{fig:prevalence}, where network adaptation takes place after
$\delta=25$~d) or termination of an outbreak as shown by the red curve
for $\delta=13$~d.  This shows that the considered adaptation scheme
can substantially limit the spreading potential of an outbreak. This
is also confirmed by the mean-field approximation
(Section~\ref{sec:mean-field}).

\begin{figure}[t!]
\includegraphics[width=0.6\textwidth]{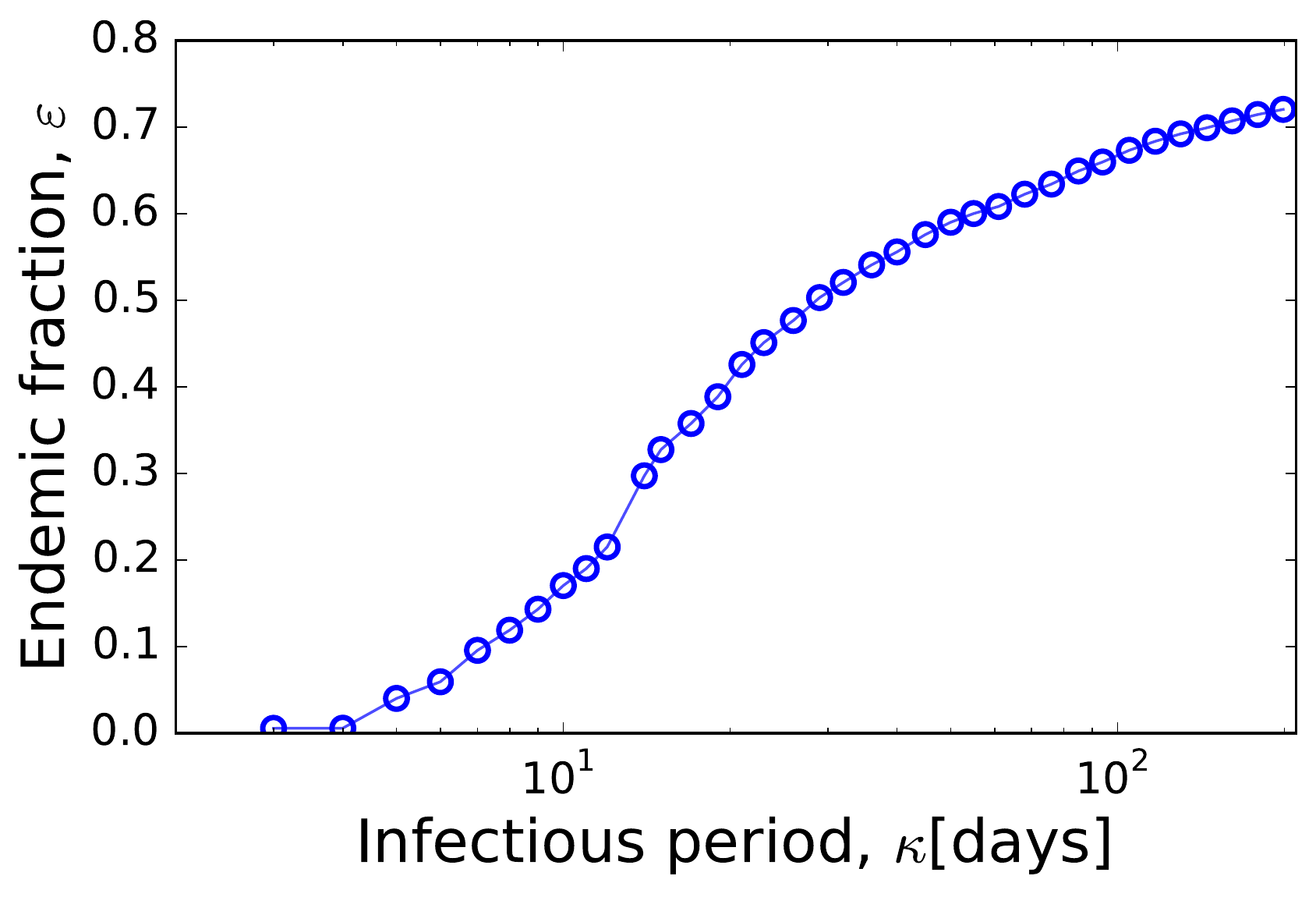}
\caption{(Color online) Endemic fraction $\varepsilon$ in dependence
  on the infectious time $\kappa$ without control adaptive rewiring.
\label{fig:endemic_fraction}}
\end{figure}

In general, different nodes as origins of infection lead to different
outcomes, some of them lead to epidemics and some do
not. Note that we always start our simulation with just one origin (or
index) node infected. In order to evaluate the influence of every
possible node as an origin of infection, we scan the whole network by
separately considering each node as the origin of an outbreak upon the
node's first appearance in the dataset. Technically we define
epidemics as persistent if after 700 days (at the end of the maximal
observation time) there still exists a non-zero prevalence in the
system.  Therefore we define the {\it endemic fraction} or the
probability of an epidemic to be sustained as the fraction of index
nodes leading to persistent epidemics
\begin{equation}
\varepsilon = \frac{N_{\text{endemic}}}{N_{\text{total}}},
\label{epsilon}
\end{equation}
where $N_{\text{total}}$ is the total number of origin nodes and and
$N_{\text{endemic}}$ is the number of origin nodes leading to
persistent epidemics.  Without adaptation, $\varepsilon$ increases
with infectious period $\kappa$
(Fig.~\ref{fig:endemic_fraction}). Only for very small infectious
periods ($\kappa\leq3$~d) there is an almost vanishing endemic fraction
due to the low frequency of network contacts. Values larger than
$10$\% are found for infectious periods larger than $5$~d. The highest
endemic fraction in our system is around $70\%$ for
$\kappa=200$~d. The considered $\kappa$-values fall in the
biologically plausible range including bacterial diseases such as
hemorrhagic diarrhea caused by {\it E.~Coli} with animals carrying the
disease up to 2 months~\cite{COR04}.
%
\begin{figure}[t!]
\includegraphics[width=0.6\textwidth]{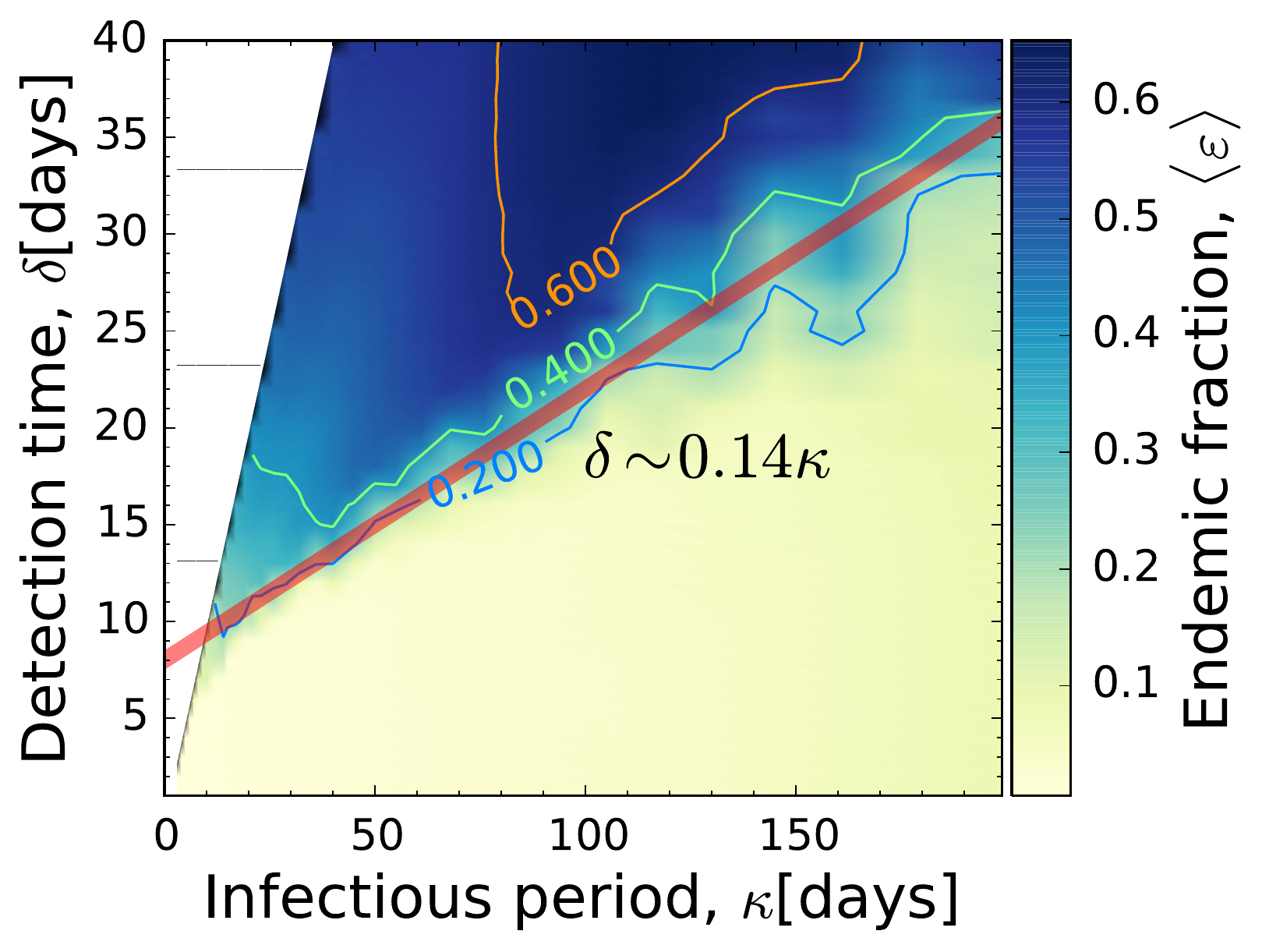}
\caption{(Color online) Dependence of the endemic fraction
  $\varepsilon$ on the infectious period $\kappa$ and detection time
  $\delta$. In the region below the red line ($\delta\sim 0.14\kappa$) the persistence fraction is less than $10\%$.
\label{fig:2d}}
\end{figure}

Besides the infectious period $\kappa$ the adaptation introduces an additional time
scale --- the detection time $\delta$. The dependence of the endemic
fraction $\epsilon$ in $(\kappa, \delta)$-parameter space is presented
in Fig.~\ref{fig:2d}. The uncontrolled case of
Fig.~\ref{fig:endemic_fraction} can be retrieved for
$\delta=\kappa$. We find that the (almost) disease-free region of
small $\varepsilon$ becomes significantly larger than in the
uncontrolled case, where the vanishing prevalence was observed for small
$\kappa\leq3$. Due to the adaptation, the persistence fraction remains
less than $10\%$ for parameter values below the red line
$\delta\approx 0.14\kappa+\text{const}$.

The behavior of the persistence fraction is also shown in Figs.~7(a)
and (b) (Supplementary Material), which shows the persistence fraction
$\varepsilon$ for fixed $\delta$ or $\kappa$, i.~e., horizontal or
vertical sections through Fig.~\ref{fig:2d}, respectively. In
Fig.~7(a) (Supplementary Material) we observe that in most cases for a
fixed detection time the persistence fraction, starting from high
values, increases intitially to eventually decrease with the
infectious period $\kappa$. In Fig.~7(b) we see that for fixed
infectious period $\kappa$ the endemic fraction monotonically
increases with the detection time $\delta$.

For recurrent epidemics like an SIS process, the impact of the
mitigation strategy by rewiring strongly depends on the node, from
which the outbreaks originates, i.e. the index node.  This
heterogeneity of the prevalence is the direct implication of the
heterogeneity in the out-components (cf. Fig~2, Supplementary
Material). The existence of clusters of nodes with similar prevalence
values and adaptation properties is consistent with the existence of
clusters of nodes with similar invasion routes as reported in
Ref.~\cite{BAJ12}.  There are subsets of index nodes, belonging to the
same cluster, which are most responsive to adaptation. This knowledge
can be exploited to target those nodes with high priority, if
resources for disease control are sparse.  To quantify the effect of
adaptation on prevalence reduction for persistent epidemics, we define the {\it
  efficacy} as
\begin{equation}
\gamma=\frac {\left(I_{\infty}-I^{*}_{\infty}\right)}{I_{\infty}},\label{gamma}
\end{equation}
where $I_{\infty}$ and $I^{*}_{\infty}$ denote the prevalence in
the unadapted and adapted cases, respectively.  Figure~5
(Supplementary Material) shows the influence of the index nodes and
their impact on the success of the control measured in terms of the
efficacy $\gamma$ for a fixed infectious period $\kappa = 28$~d and
different detection times $\delta$. For small $\delta$, we find the
efficacy peaked at 90\%. As the detection time becomes larger, the
control becomes less effective indicated by smaller prevalence
reduction. The mean efficacy averaged over all nodes as starting
points is presented in Fig~8 in Supplementary Material. To
characterize the heterogeneity of the distribution of the $\gamma$
values shown in Fig.~5 (Supplementary Material), we also compute its
entropy. Its dependence on the infectious period and detection time is
presented in Fig.~9 (Supplementary Material).
\section{Mean-field Description}\label{sec:mean-field}
We approximate the deterministic dynamics on a temporal network
described in Section~\ref{sec:methods}(See Fig.~1) with the following system of
stochastic reactions
\begin{eqnarray*}
S+I & \stackrel{\alpha}{\rightarrow} & 2I\\
S+D & \stackrel{\alpha^{\prime}}{\rightarrow} & 2I\\
I & \stackrel{\nu}{\rightarrow} & D\\
D & \stackrel{\mu}{\rightarrow} & S.
\end{eqnarray*}
The first reaction describes the usual infection of a susceptible by
an infective with the infection rate $\alpha$ which in the
deterministic case could correspond to the average daily out-degree of a
node $\alpha\sim\langle k_{\rm out}\rangle$. The second reaction corresponds to
the effective force of infection due to the occasional rewiring to
infected nodes (See Fig.~1, link {$\bf c$}) which reads $\alpha^{\prime}SD$
with the effective infection rate $\alpha^{\prime}=\alpha
I/(I+S)$. The third reaction represent the detection with the rate
$\nu\sim1/\delta$. And the last one the recovery with the rate
$\mu\sim 1/(\kappa-\delta)$.
Using the fractions $j=I/N$, $s=S/N,$ and
$z=D/N$, the corresponding set of differential equations reads
\begin{eqnarray*}
\frac{dj}{dt} & = & \alpha\left(1+\frac{z}{j+s}\right)js-\nu j,\\
\frac{ds}{dt} & = & -\alpha\left(1+\frac{z}{j+s}\right)js+\mu z,\\
\frac{dz}{dt} & = & \nu j-\mu z.
\end{eqnarray*}
Note that the total number of nodes is conserved $j+s+z=1$. Thus we
could eliminate the third equation.  Usually, to find the
deterministic threshold for a disease outbreak, stability of the
disease-free fix point $(j=0,s=1)$ is considered.  The eigenvalues are
\[
\lambda_1=-\mu=\frac{1}{\delta-\kappa}
\]
and 
\[
\lambda_2=\alpha-\nu=\alpha-1/\delta.
\]
Because $\delta<\kappa$, the first eigenvalue is always negative. From
the outbreak condition $\lambda_2>0$, we have $\alpha\delta>1$, i.e.\ is the standard threshold
condition, without any dependence on the infectious period $\kappa$. 
It seems that this mean-field approximation does not reproduce the observed threshold
$\delta\sim \text{const}\times\kappa+\text{const}$ correctly (see Fig.~4).
The endemic prevalence is given by 
\[
j_{\infty}^{*}=\left(\frac{\nu}{\mu}+\frac{\alpha}{\alpha-\nu}\right)^{-1}.
\]
Note that in the limit $\mu\rightarrow\infty$ or equivalently $\delta=\kappa$,
corresponding to the unadapted case we recover the well-known result
for an SIS model: $j=(\alpha-\nu)/\alpha$. Thus the efficacy (\ref{gamma}) reads
\[
\gamma = \left(\frac{\nu}{\alpha}+\frac{\mu}{\alpha-\nu}\right)^{-1}.
\]
The endemic values of $s^{*}$ and $z^{*}$ variables
are given by the expressions
\begin{eqnarray*}
s_{\infty}^{*} & = & \frac{\nu}{\alpha-\nu}j_{\infty}^{*},\\
z_{\infty}^{*} & = & 1-j_{\infty}^{*}-s_{\infty}^{*}.
\end{eqnarray*}
\section{Conclusions}
\label{sec:discussion}
We have investigated disease control based on an adaptive rewiring
strategy of a time-varying network to mitigate the effect of a
recurrent deterministic epidemic. This control measure relies on
isolating infectious nodes and thus is different from most of control
approaches proposed for temporal
networks~\cite{FAS15a,LEE12,STA13,BUE13}. We have considered a
SIS-type dynamics on the nodes and introduced a detection time, after
which links can be rewired to isolate infectious nodes. As an
exemplary temporal contact network with real-world application, we
analyzed an animal trade network, where each trading event corresponds
to a contact between two agricultural premises. The network of farms
can be seen as a contact network with nodes in a susceptible or an
infected state.

We have found that for recurrent epidemics, the starting point of an
outbreak is very important for the course of the epidemics: it either
dies out or becomes endemic with different prevalence levels. This
happens due to the heterogeneity of the subset of the network
reachable from the specific first (index) node. Accordingly, we have
found that the impact of a mitigation strategy by network adaptation
is similarly variable. The region of disease parameters, where most of
the index nodes lead to vanishing epidemics, can be substantially
extended using the proposed adaptive rewiring strategy. To effectively
control the epidemic, the detection times should be less than $10$
days. Moreover, there is a range of detection time values between 7
and 10 days, which lead to especially effective mitigation of
epidemics with an infectious period around 30 days. This might be due
to the interplay of the internal time scales of the system. We have
shown that the success of an adaptation depends also on the parameters
of the epidemics and, for instance, saturates for very long infectious
periods.

In the presented work, we have provided a proof of concept and
reported on the effect of modification of the contact network. The
model can be further detailed and extended following a
metapopulational approach, which takes into account the number of
animals traded or present in the premises, heterogeneity of
parameters, as well as stochastic effects along the lines of, for
instance, Ref.~\cite{GRI05,LAM15a,BIG07,AUD01,ORT06a}. This, however,
is beyond the scope of this study, but a promising topic for the
future.

\section*{Data accessibility}
The dataset on animal mobility due to the trade between different
agricultural holdings used in our study, was extracted from the
database on pig trade in Germany~\textit{HI-Tier} \footnote{The
  HI-Tier database (https://www.hi-tier.de) is administered by the
  Bavarian State Ministry for Agriculture and Forestry on behalf of
  the German federal states} established according to EU
legislation \footnote{Directive 2000/15/EC of the European Parliament
  and the Council of 10 April 2000 amending Council Directive
  64/432/EEC on health problems affecting intra-community trade in
  bovine animals and swine.}. Due to privacy reasons the data is
available to designated German authorities.
\section*{Competing interests}
The authors declare no competing interests.
\section*{Authors' contributions}
VB and PH designed the study. HL prepared the empirical data. VB, AF,
and FF implemented the model and analyzed the data. VB and PH were the
lead writers of the manuscript. All authors gave final approval for
publication.

\section*{Acknowledgements}
The authors would like to thank Thilo Gross and Thomas Selhorst for
fruitful discussions.
\section*{Funding}
VB and PH acknowledge support by Deutsche Forschungsgemeinschaft in
the framework of Collaborative Research Center 910.
\bibliographystyle{prsty-fullauthor}
\bibliography{export}
\clearpage
\setcounter{figure}{0}
\section*{Supplementary Material}
\begin{figure}[ht!]
\includegraphics[width=0.60\textwidth]{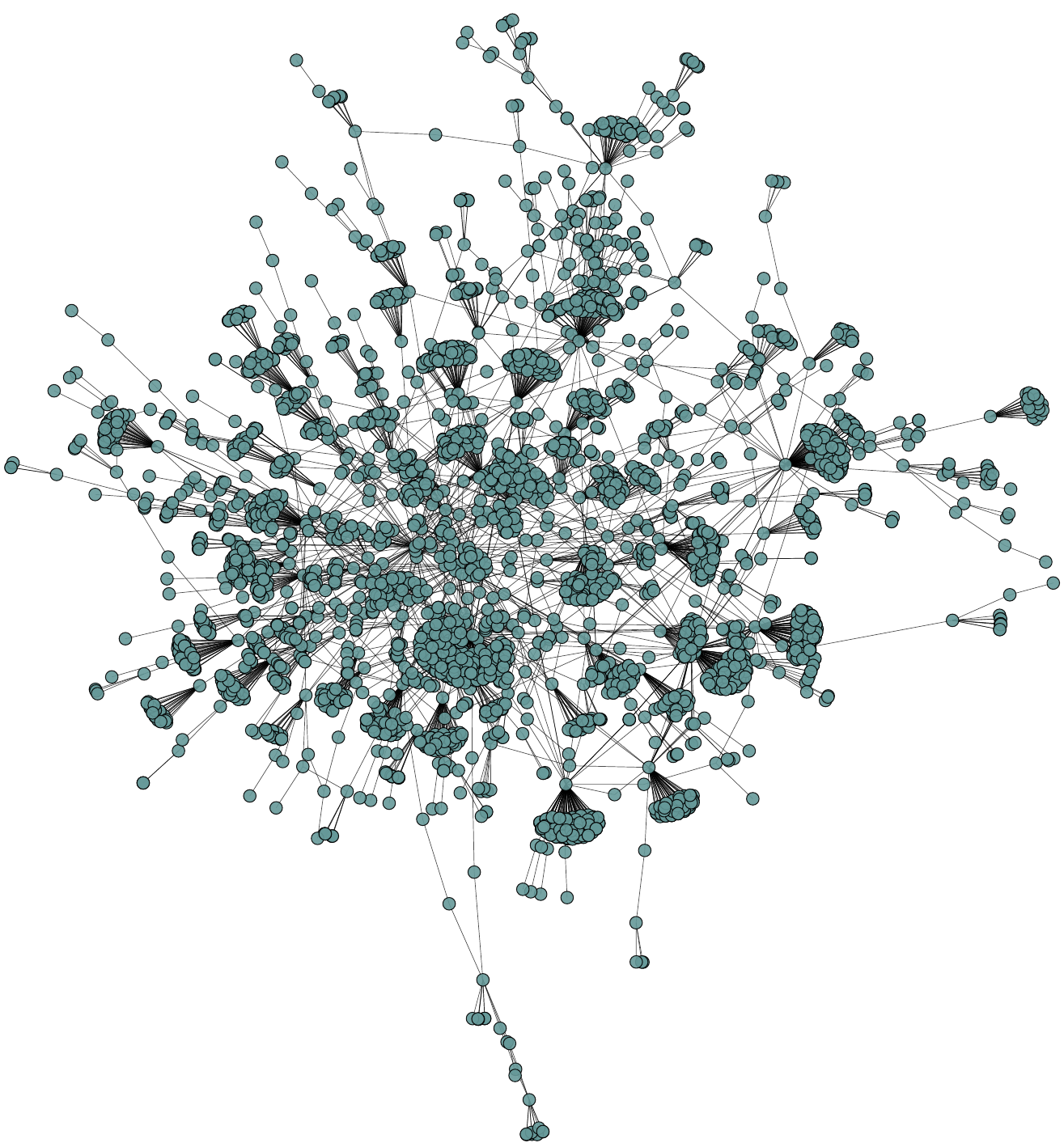}
\caption{Visualization of the aggregated network of trading
  contacts, appearing at least $50$ times in the dataset. Here the
  links are considered as undirected.
\label{fig:net_viz}}
\end{figure}

The network of animal trade is visualized in
Fig.~\ref{fig:net_viz}. The network exhibits a tree-like structure and
can be interpreted as a hierarchical supply-chain network. This is of
particular interest for the spread of diseases, because outbreaks will
have different impact depending on where they first occur.
\begin{table}[htp]
\caption{Basic network properties of en exerpt of the time-aggregated German
  pig-trade network. Properties denoted by asterisk are for the
  largest connected component of the network considered as
  undirected.}
\begin{center}
\begin{tabular}{lr}
\toprule
\textbf{Property} & \textbf{Value}\\
\hline
Number of nodes & 15,569\\
Number of edges & 748,430\\
\hline
Average daily number of nodes (except Sundays) & 1220 \\
Average daily number of edges (except Sundays) & 1141\\
Average number of nodes (working days) & 1346 \\
Average number of edges (working days) & 1258 \\
Average number of nodes (Saturdays) & 932\\
Average number of edges (Saturdays) & 844\\
Average number of nodes (Sundays) & 43 \\
Average number of edges (Sundays) & 27 \\
\hline
Diameter$^{*}$& 13\\
Average shortest path length$^{*}$ & 4.081\\
Average clustering coefficient$^{*}$ & 0.1779\\
\hline
\hline
\end{tabular}
\end{center}
\label{tab:properties}
\end{table}%

\begin{figure}[t!]
 \includegraphics[width=0.5\textwidth]{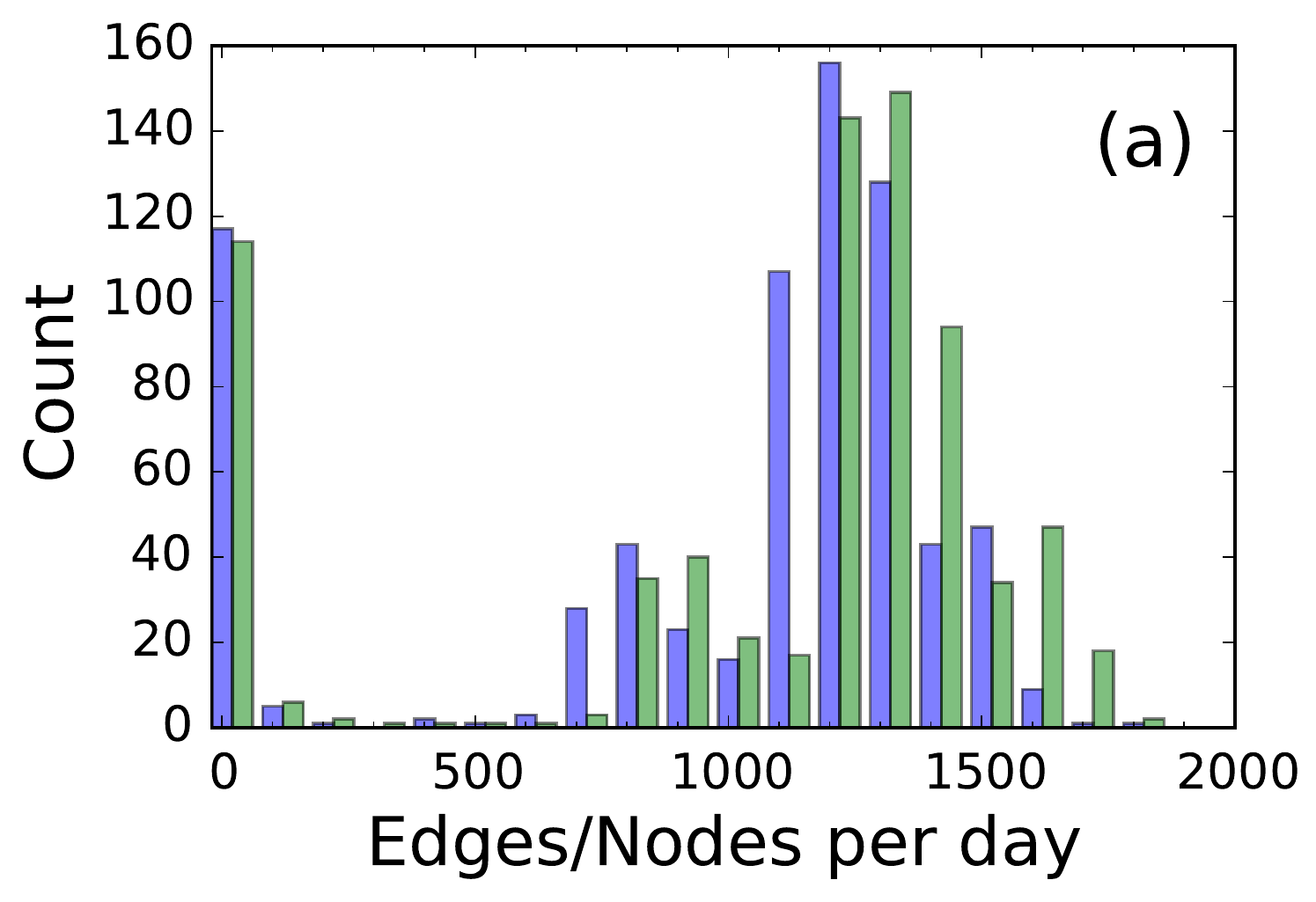}%
 \includegraphics[width=0.5\textwidth]{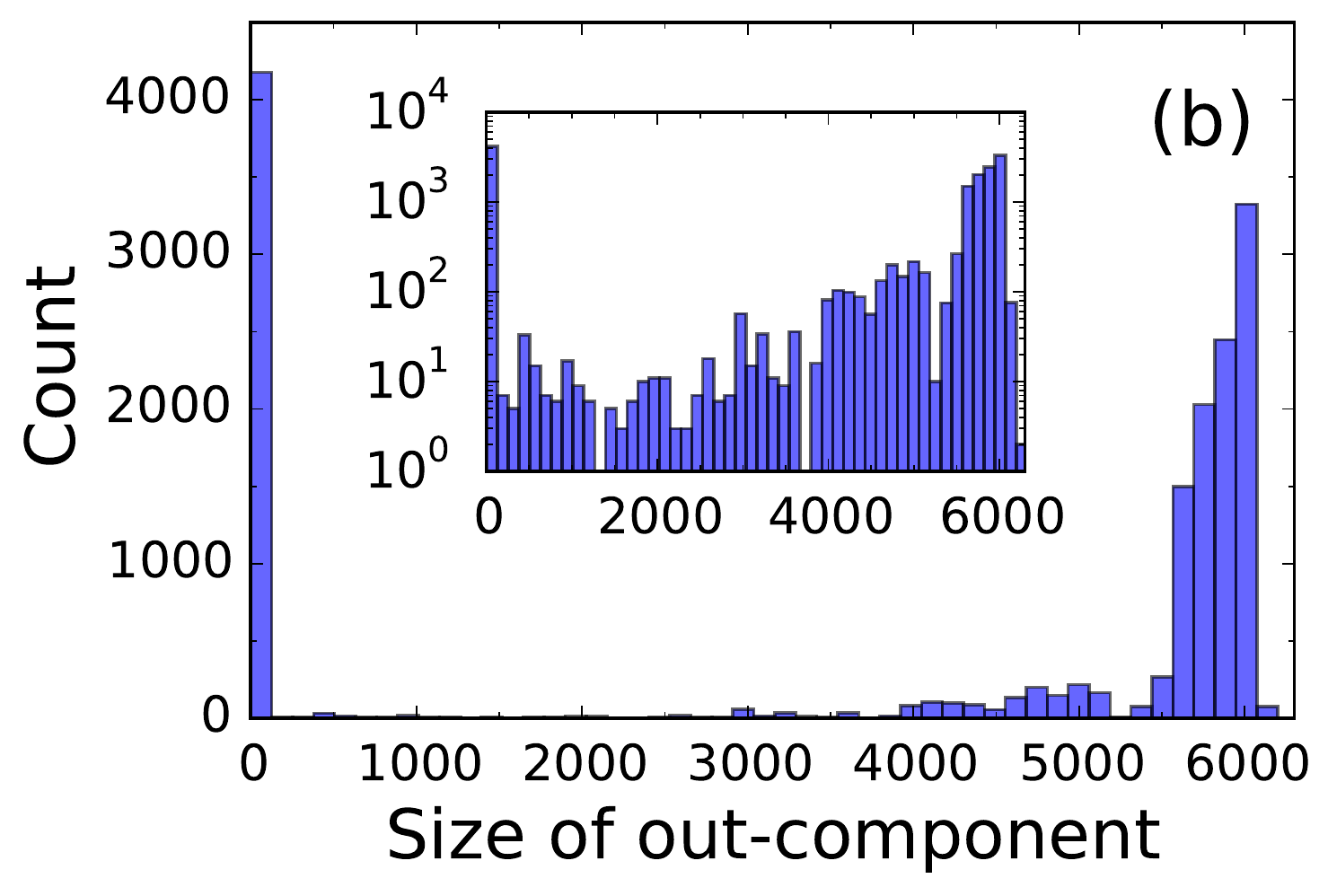}
\caption{ (a) Histogram of the number of active edges (blue) and nodes
  (green) (daily resolution). (b) Distribution of the size of
  out-components for all nodes. The inset shows the same on a semi-log
  scale.%
\label{fig:Snapshot size and out components}}
\end{figure}
\begin{figure}[ht!]
\includegraphics[width=\textwidth]{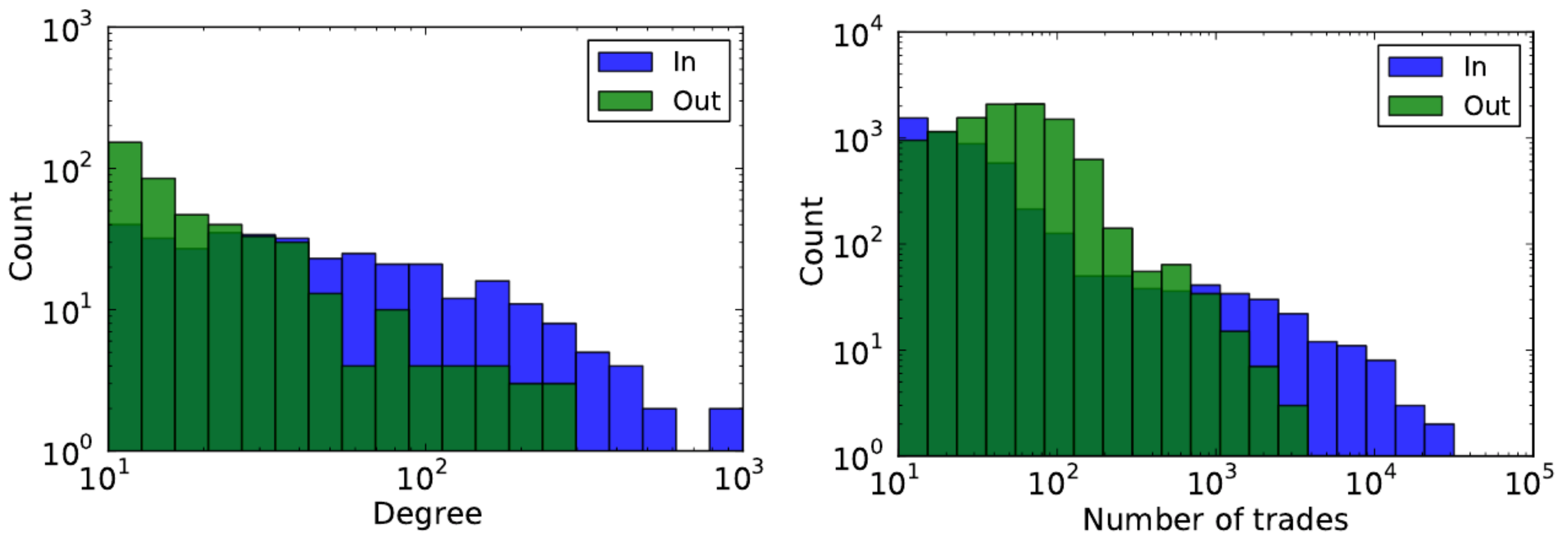}
\caption{Histogram of node degree (blue: in-degree, green: out-degree)
 and activity (blue: in-coming activity, green: out-going activity) of the time-aggregated graph.\label{fig:degree and activity}}
\end{figure}
\begin{figure}[ht!]
 \includegraphics[width=0.5\textwidth]{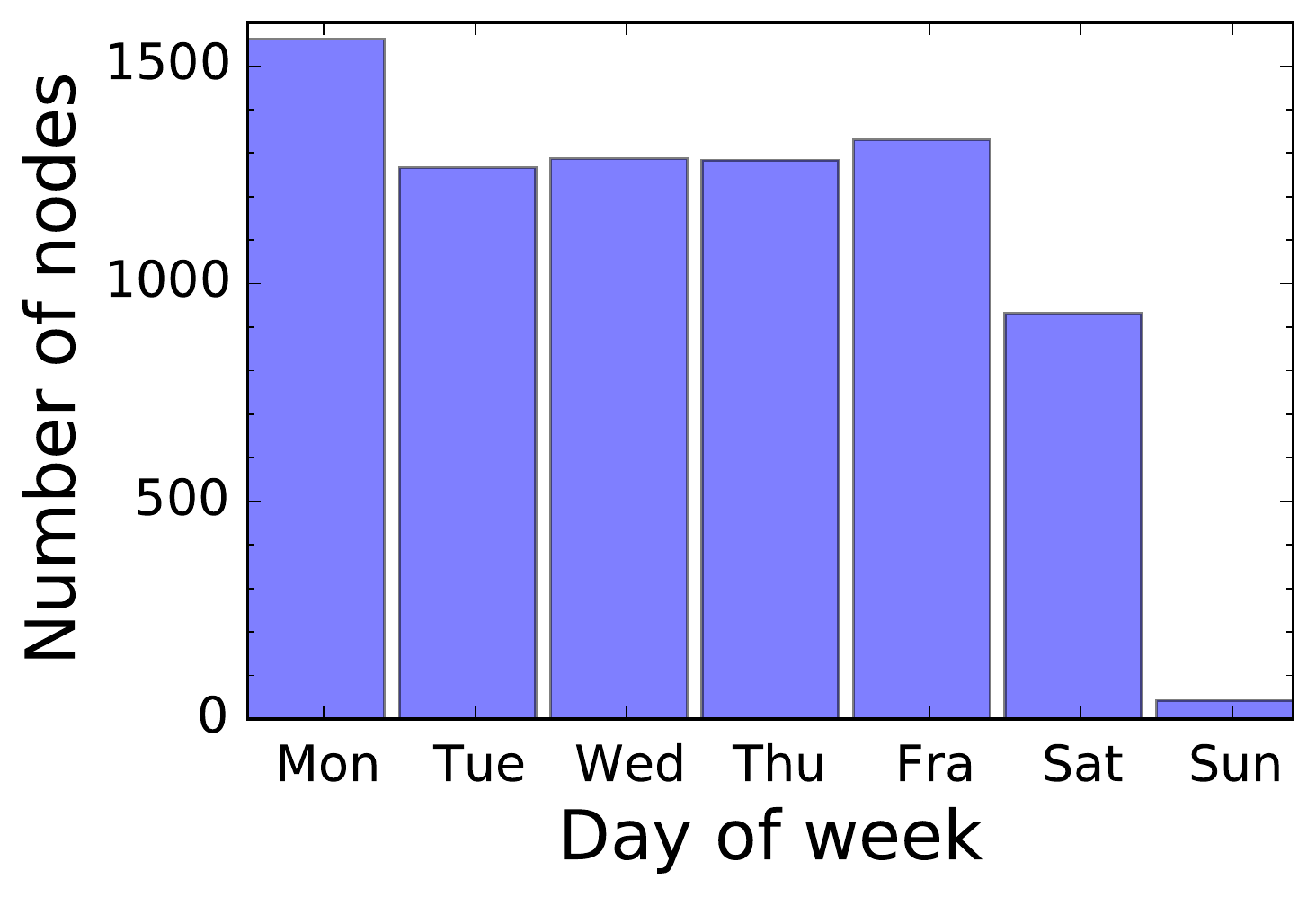}%
 \includegraphics[width=0.5\textwidth]{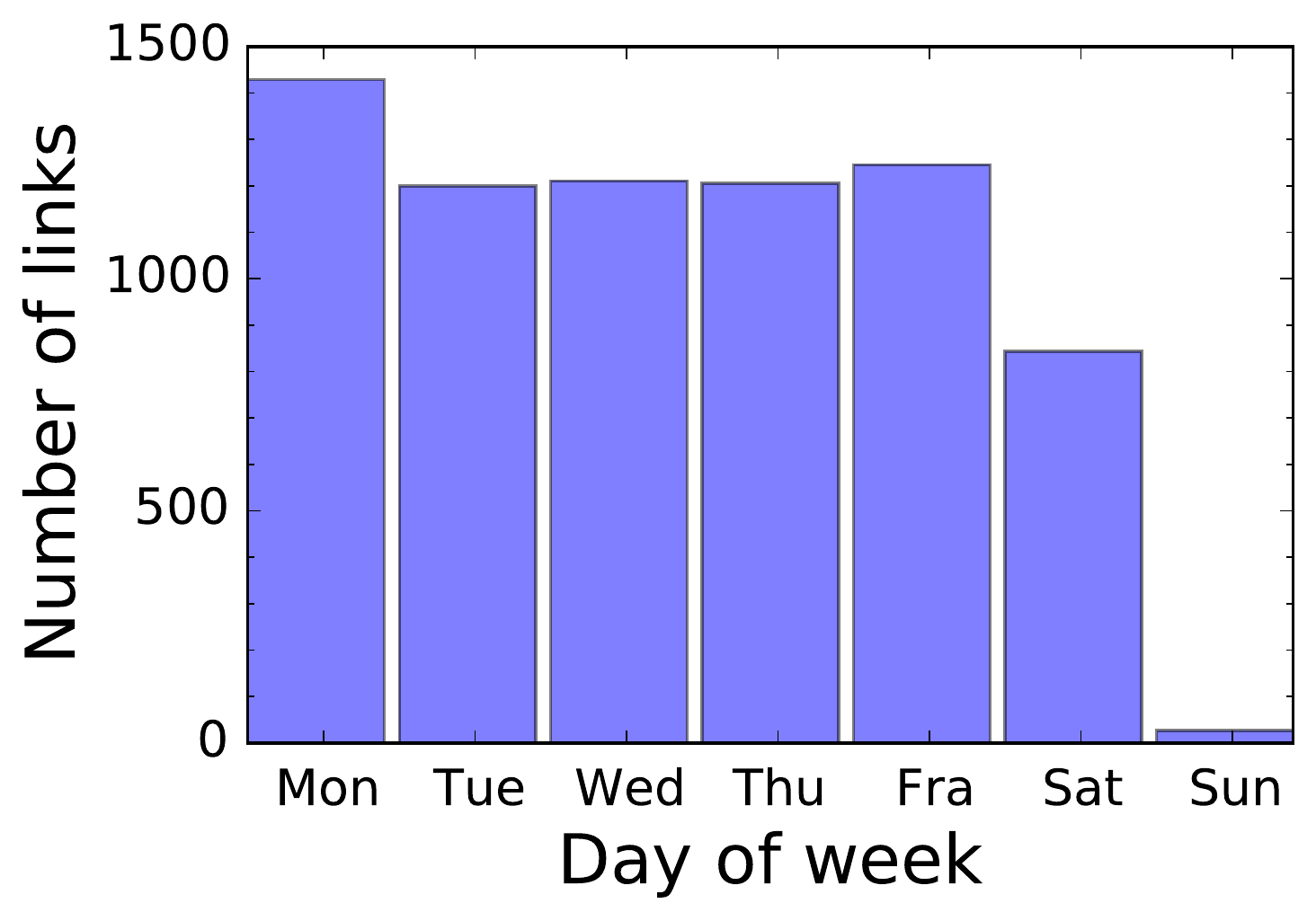}
\caption{ (Average number of links and nodes on different week days.
\label{fig:nodes and links}}
\end{figure}

In addition to Tab.~\ref{tab:properties}, basic network
characteristics are presented in Figs.~\ref{fig:degree and activity}
and \ref{fig:nodes and links}.  Distribution of both the node degrees
and the node activity aggregated over the whole period of observation
is very broad as shown in Fig.~\ref{fig:degree and activity}.

Figure~\ref{fig:nodes and links} depicts the average number of nodes
and links resolved for each day of the week. Mondays show the highest
numbers and other working days a similar level. On Saturdays, the
network is less dense and the numbers of nodes and links are the lowest
for Sundays.
\begin{figure}[ht!]
\includegraphics[width=0.33\textwidth]{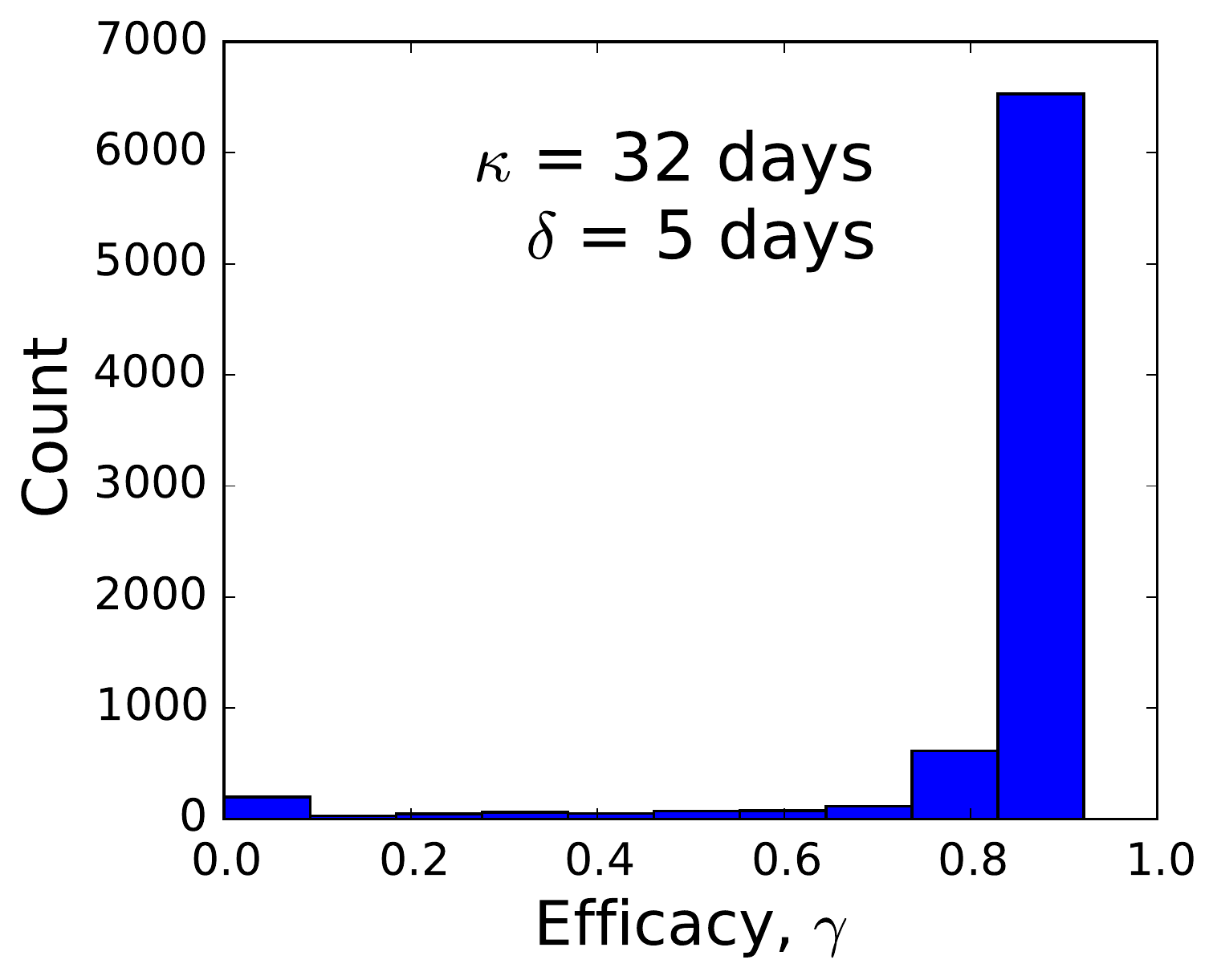}%
\includegraphics[width=0.33\textwidth]{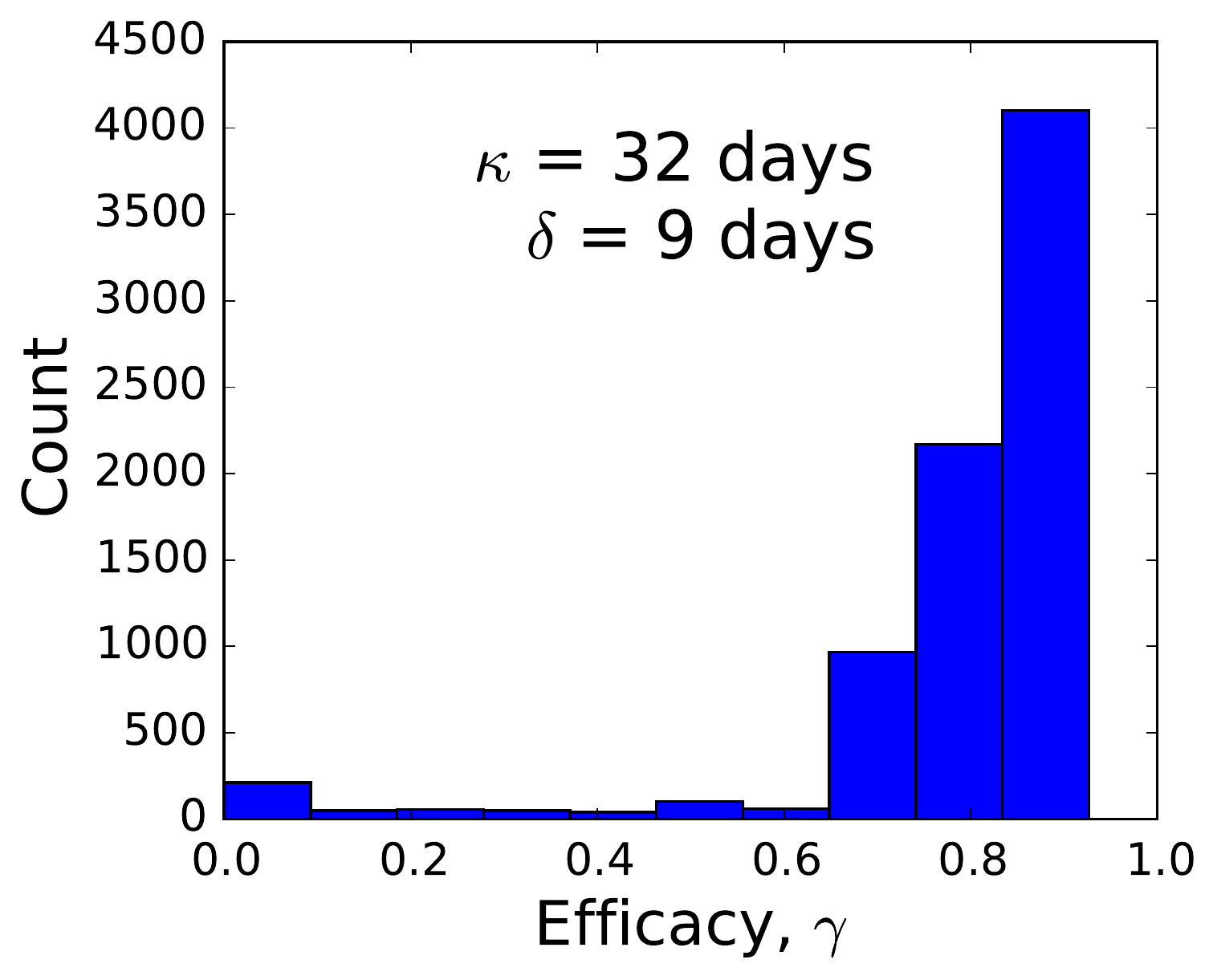}%
\includegraphics[width=0.33\textwidth]{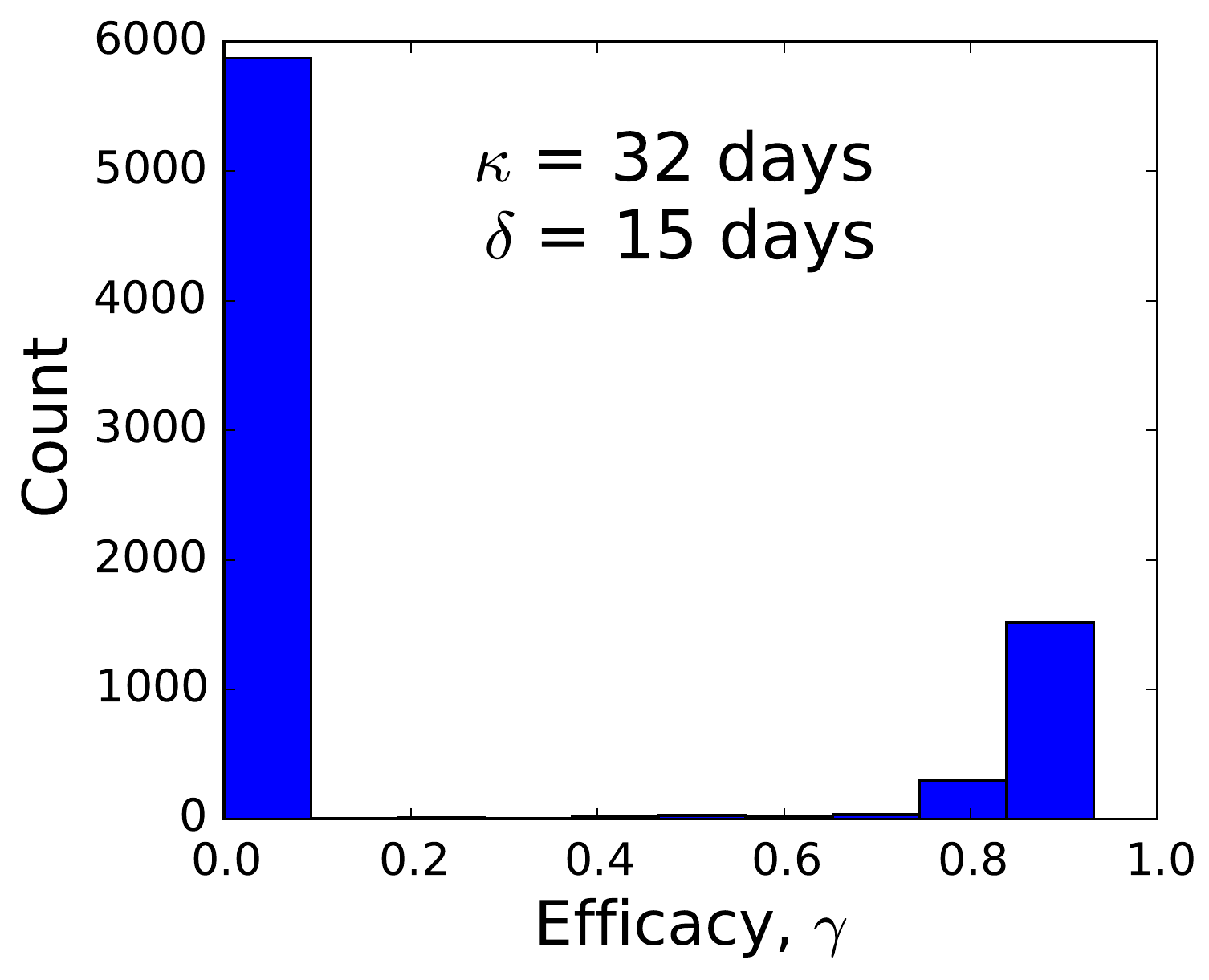}%
\caption{Histogram of the prevalence reduction $\gamma$ for infectious
  period $\kappa=32$~d and different detection times $\delta$ as
  indicated in the panels. Each node in the network is chosen as a
  starting node in a separate simulation with initial infection upon
  its first appearance in the dataset.
\label{fig: prev_hist}}
\end{figure}

Figure~\ref{fig: prev_hist} shows the influence of the index nodes and
their impact on the success of the control measured in terms of the
prevalence reduction $\gamma$~(\ref{epsilon})for a fixed infectious period $\kappa=32$~d
and different detection times $\delta$. For small $\delta$, we find a
prevalence reduction peaked at 90\%. As the detection time becomes
larger, the control becomes less effective indicated by smaller
prevalence reduction. The mean prevalence reduction averaged over all
nodes as starting points is presented in Fig~\ref{fig:prev_reduction}.

\begin{figure}[t!]
\includegraphics[width=0.48\textwidth]{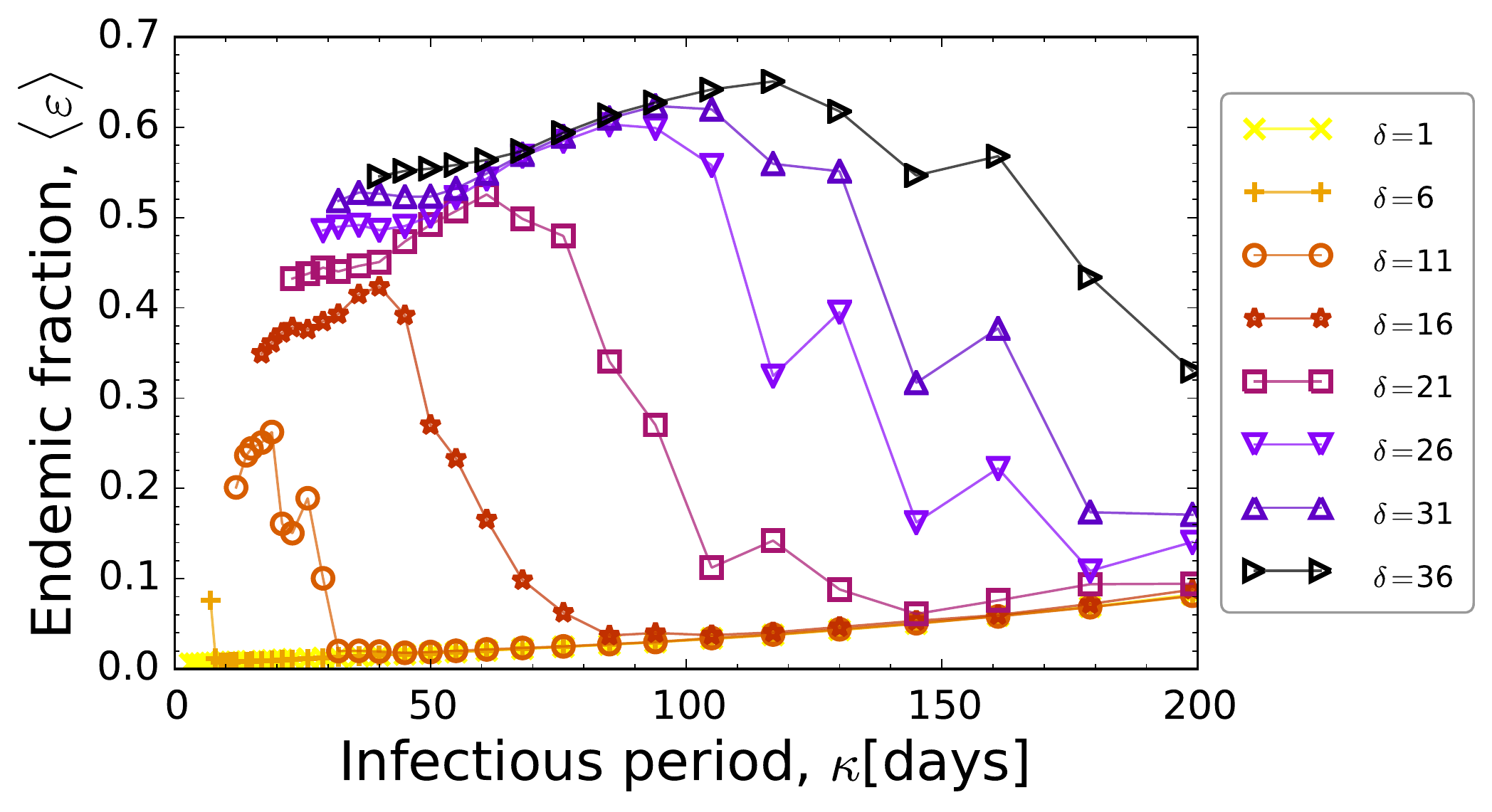}
\includegraphics[width=0.48\textwidth]{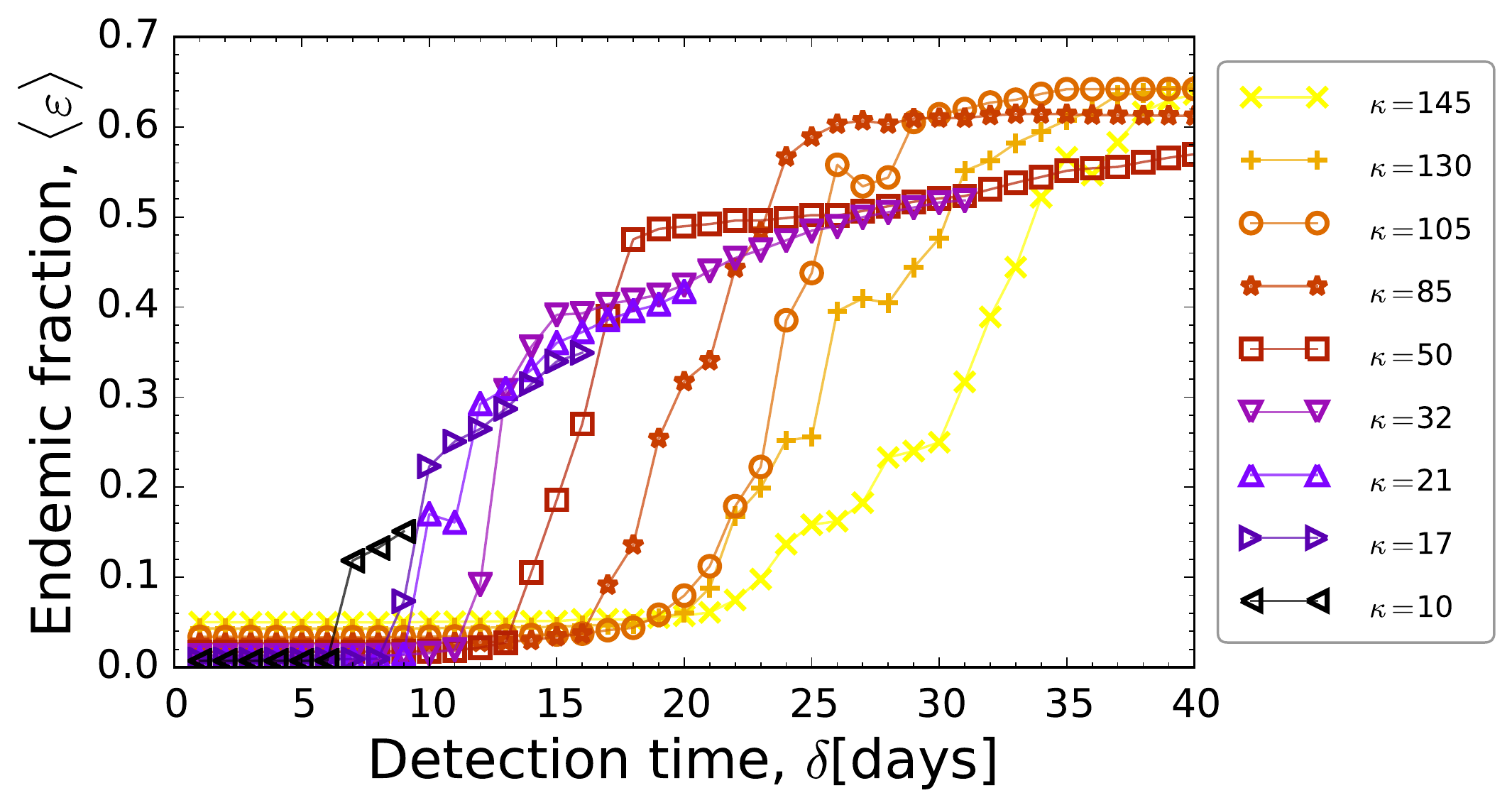}
\caption{Dependence of the endemic fraction $\varepsilon$: {\it left}
  on the infection period $\kappa$ for different detection times
  $\delta$ and {\it right} on the infectious period $\delta$ for
  different detection times $\kappa$.
\label{fig:eps_delta_kappa}}
\end{figure}
Figs.~\ref{fig:eps_delta_kappa}(a) and (b) shows the dependence of the
endemic fraction $\varepsilon$~(\ref{epsilon}) on $\kappa$ (for fixed $\delta$) and on
$\delta$ (for fixed $\kappa$) correspondingly.

\begin{figure}[ht!]
\includegraphics[width=0.60\textwidth]{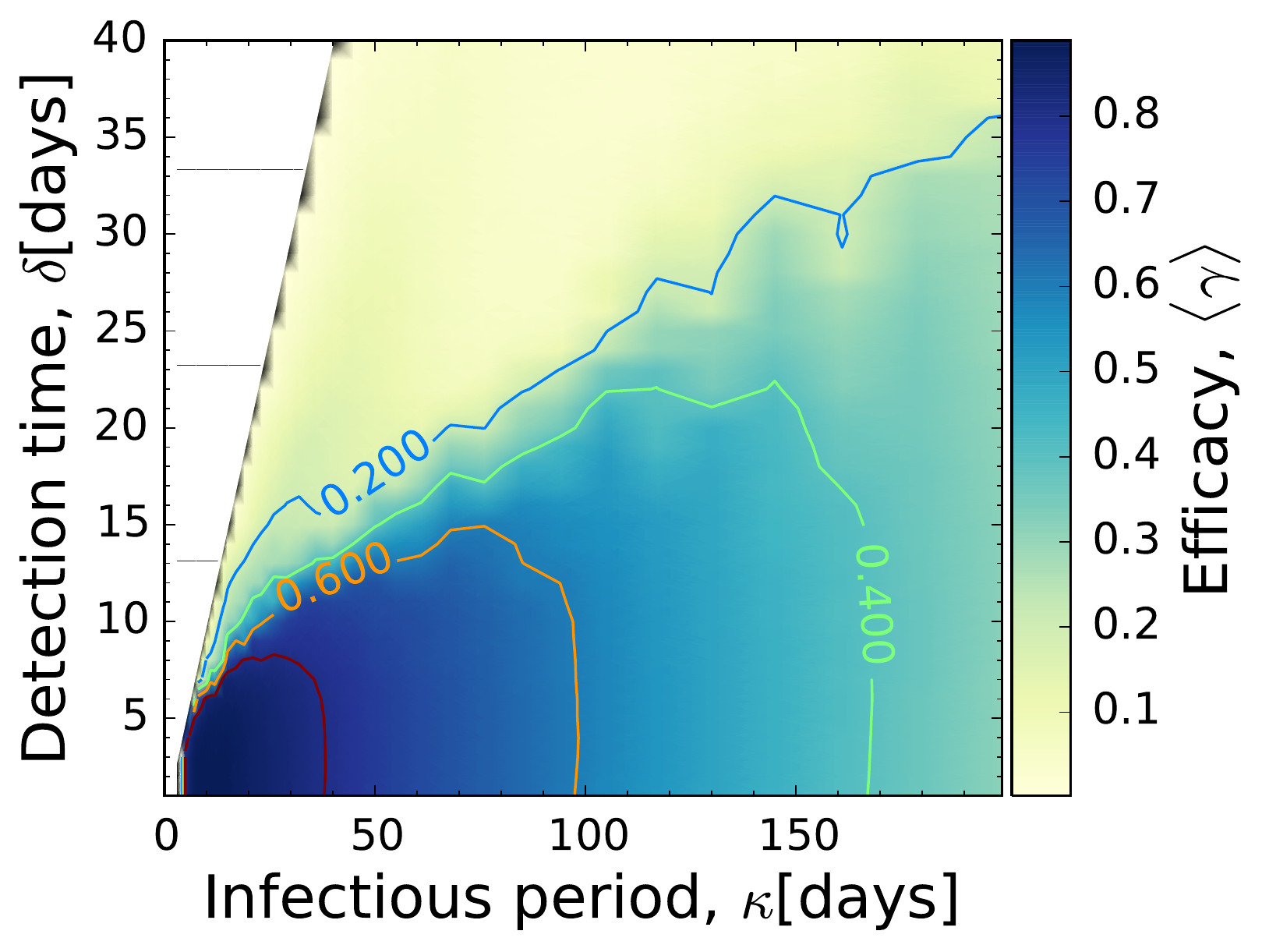}
\caption{%
Prevalence reduction $\varepsilon$ in dependence on both infectious period and detection time.
\label{fig:gamma_k_d}}
\end{figure}
\begin{figure}[ht!]
\includegraphics[width=0.5\textwidth]{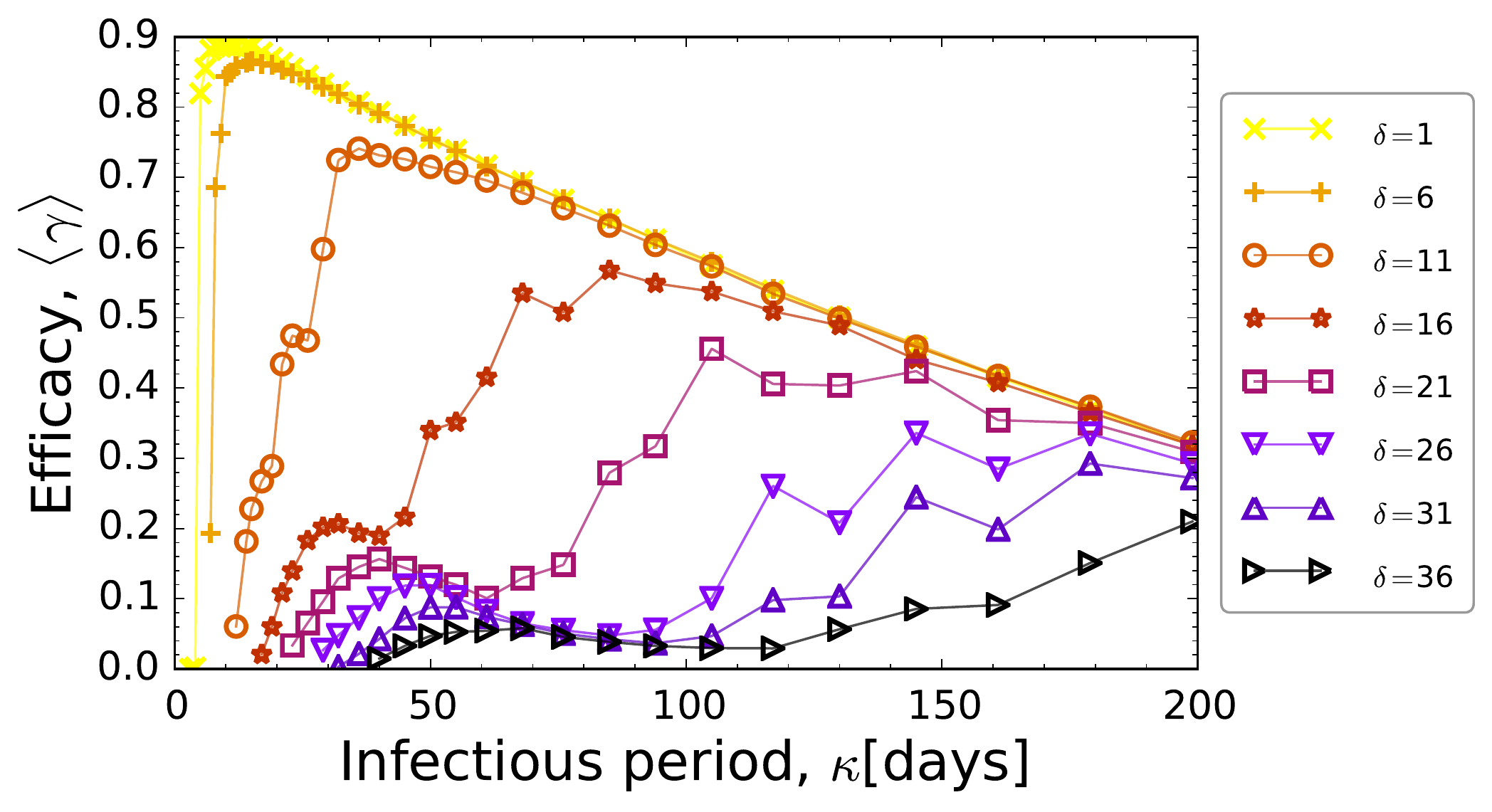}%
\includegraphics[width=0.5\textwidth]{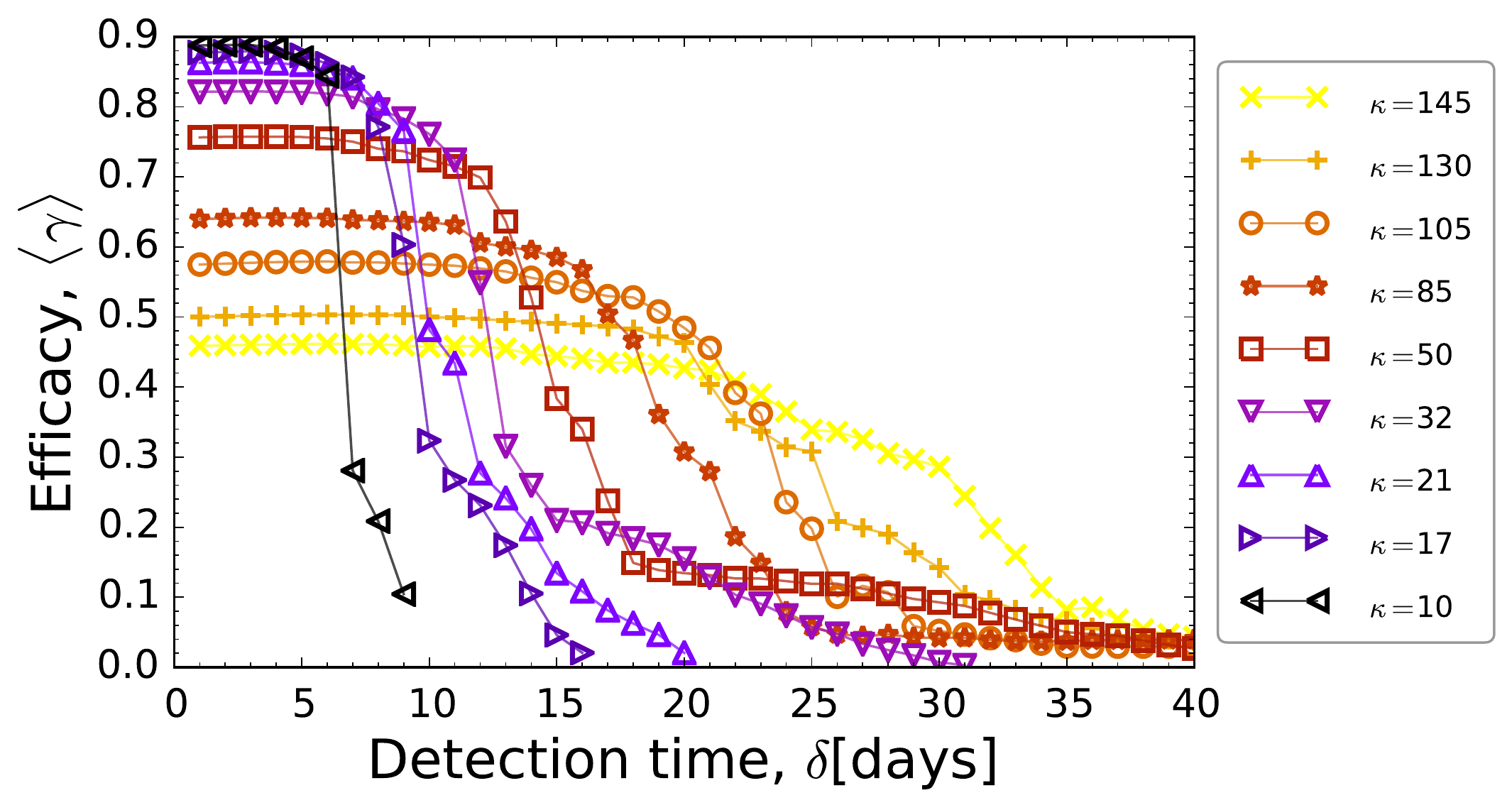}
\caption{Efficacy or relative prevalence reduction $\gamma$ averaged over all possible
  index nodes for sustained epidemics in dependence on $\kappa$ (with
  fixed $\delta$, left panel) and on $\delta$ (with fixed $\kappa$,
  right panel).}
\label{fig:prev_reduction}
\end{figure}
\begin{figure}[ht!]
\includegraphics[width=0.48\textwidth]{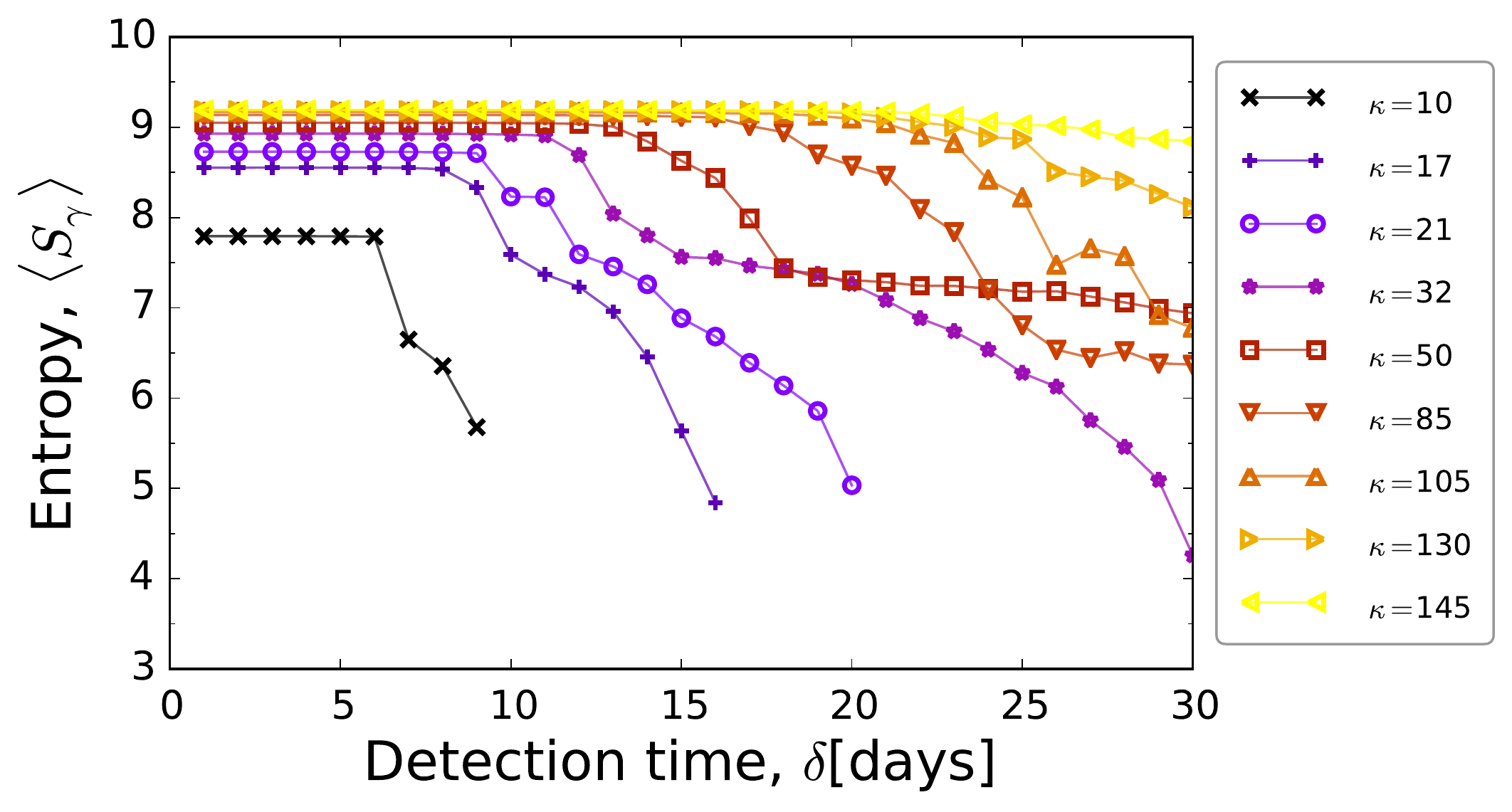}
\includegraphics[width=0.48\textwidth]{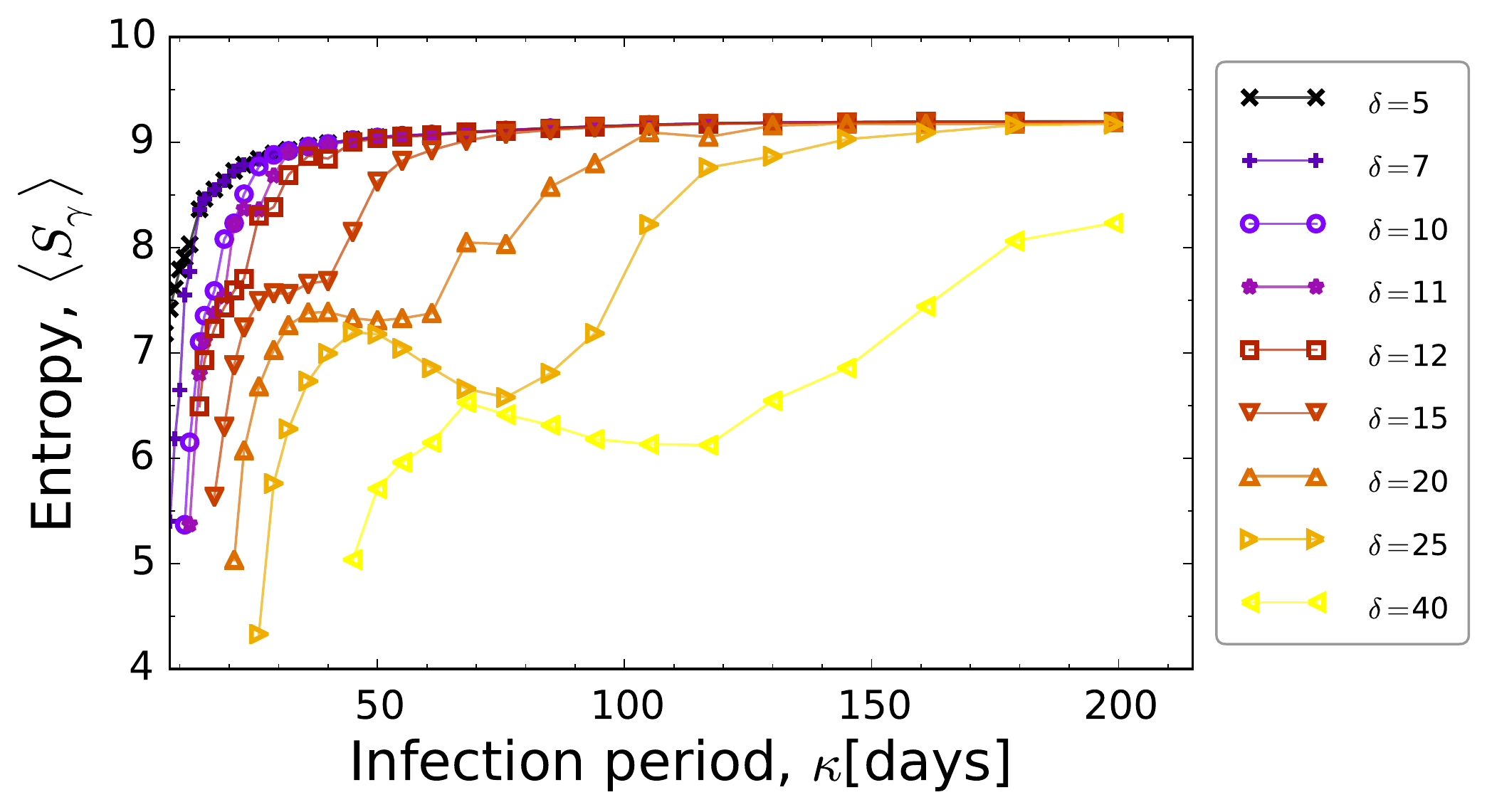}
\caption{Entropy in dependence on $\kappa$ and $\delta$, which
  characterizes the heterogeneity in the distribution of efficacy
  $\gamma$ for different index nodes.}
\label{fig:entropy}
\end{figure}
Figure~\ref{fig:prev_reduction} depicts the efficacy or prevalence
reduction for sustained epidemics $\gamma$~(\ref{gamma}) in dependence
on $\kappa$ and $\delta$. This quantity shows how useful the control
is, even if we cannot totally stop the disease, which became endemic,
and to what extent we could lower the prevalence. There is a fast
decrease in the efficacy of the control measures as the detection
takes longer.

The efficacy or prevalence reduction $\gamma$ is highly dependent on
the index node, and to characterize the heterogeneity we depict in
Figure~\ref{fig:entropy} the entropy of the efficacy or prevalence
reduction $\gamma$ defined as
$$S_\gamma = -\sum_i \pi(\gamma_i) \log[\pi(\gamma_i)],$$ where the
index $i$ enumerates all index nodes, in dependence on $\kappa$ and
$\delta$. $\pi(\gamma_i)$ is the probability for an index node $i$
to have the efficacy $\gamma_i$. Entropy in dependence on $\delta$
(Fig.~\ref{fig:entropy}, left panel) possesses clearly two flat
levels, decreasing from a high to a low one with increasing
$\delta$. Thus with early detection, different index nodes lead to
very heterogeneous prevalence levels. Later detection leads to
epidemics with similar efficacy levels. The dependence of the entropy
on the infectious period $\kappa$ (Fig.~\ref{fig:entropy}, right
panel) exhibits maxima for the intermediate values of $\kappa$. The
largest heterogeneity in the efficacy for intermediate infectious
periods might be due to the interplay of internal scales of the
temporal network.

\end{document}